\documentclass{article}
\usepackage{spconf,amsmath,graphicx}
\usepackage{enumitem}
\usepackage{float}
\setlist{nosep, leftmargin=14pt}
\usepackage{mwe} 
\usepackage{array}
\usepackage{booktabs}
\usepackage{bm}

\DeclareMathOperator*{\argmin}{arg\,min} % Jan Hlavacek

\usepackage{microtype}
\usepackage[square, sort, comma, numbers, compress]{natbib}

\begin{document}

% Title.
\title{Label Propagation via Random Walk for Training Robust Thalamus Nuclei Parcellation Model from Noisy Annotations}

% Single address.
% \name{
% \text{Anqi Feng}$^{1}$\qquad
% \text{Yuan Xue}$^{2}$\qquad 
% \text{Yuli Wang}$^{1}$\qquad 
% \text{Chang Yan}$^{3}$\qquad
% \textit{Zhangxing Bian}$^{2}$\qquad\\
% \textit{Muhao Shao}$^{2}$\qquad
% \textit{Jiachen zhuo}$^{4}$\qquad
% \textit{Aaron Carass}$^{2}$\qquad 
% \textit{Jerry L. Prince}$^{2}$}

\address{
$^{1}$ Department of Biomedical Engineering, Johns Hopkins School of Medicine, USA\\
$^{2}$ Department of Electrical and Computer Engineering, Johns Hopkins University, USA\\
$^{3}$ Department of Information Technology and Electrical Engineering, ETH Zurich, Switzerland \\
$^{4}$ Department of Diagnostic Radiology and Nuclear Medicine,\\ University of Maryland School of Medicine, USA}

\maketitle
%
%
% -------------------Abstract ----------------------
\begin{abstract}
Data-driven thalamic nuclei parcellation depends on high-quality manual annotations.
However, the small size and low contrast changes among thalamic nuclei, yield annotations that are often incomplete, noisy, or ambiguously labelled.
To train a robust thalamic nuclei parcellation model with noisy annotations, we propose a label propagation algorithm based on random walker to refine the annotations before model training.
A two-step model was trained to generate first the whole thalamus and then the nuclei masks.
We conducted experiments on a mild traumatic brain injury~(mTBI) dataset with noisy thalamic nuclei annotations.
Our model outperforms current state-of-the-art thalamic nuclei parcellations by a clear margin. We believe our  method can also facilitate the training of other parcellation models with noisy labels.
\end{abstract}
%
%
% ---------------------Introduction--------------------
\section{Introduction}
\label{sec:intro}
The thalamus is an essential relay station in the brain that passes information from peripheral nerves to the cortex~\cite{sherman2006exploring} and also reflects multiple behavioral states~\cite{sherman2002role}. The thalamus has a complicated structure, consisting of multiple functional clusters called \textit{nuclei}. These nuclei serve a wide variety of neurological functions and are closely associated with many neural disorders such as schizophrenia, multiple sclerosis~\cite{glaister2017ni}, Alzheimer's disease, epilepsy, Huntington's disease, and dyslexia~\cite{iglesias2018probabilistic} where both atrophy and functional changes can be observed. Therefore, accurate segmentation of the thalamus and thalamic nuclei can be helpful in clinical medicine.

\begin{figure}[!t]
\centering
\includegraphics[width=1\columnwidth]{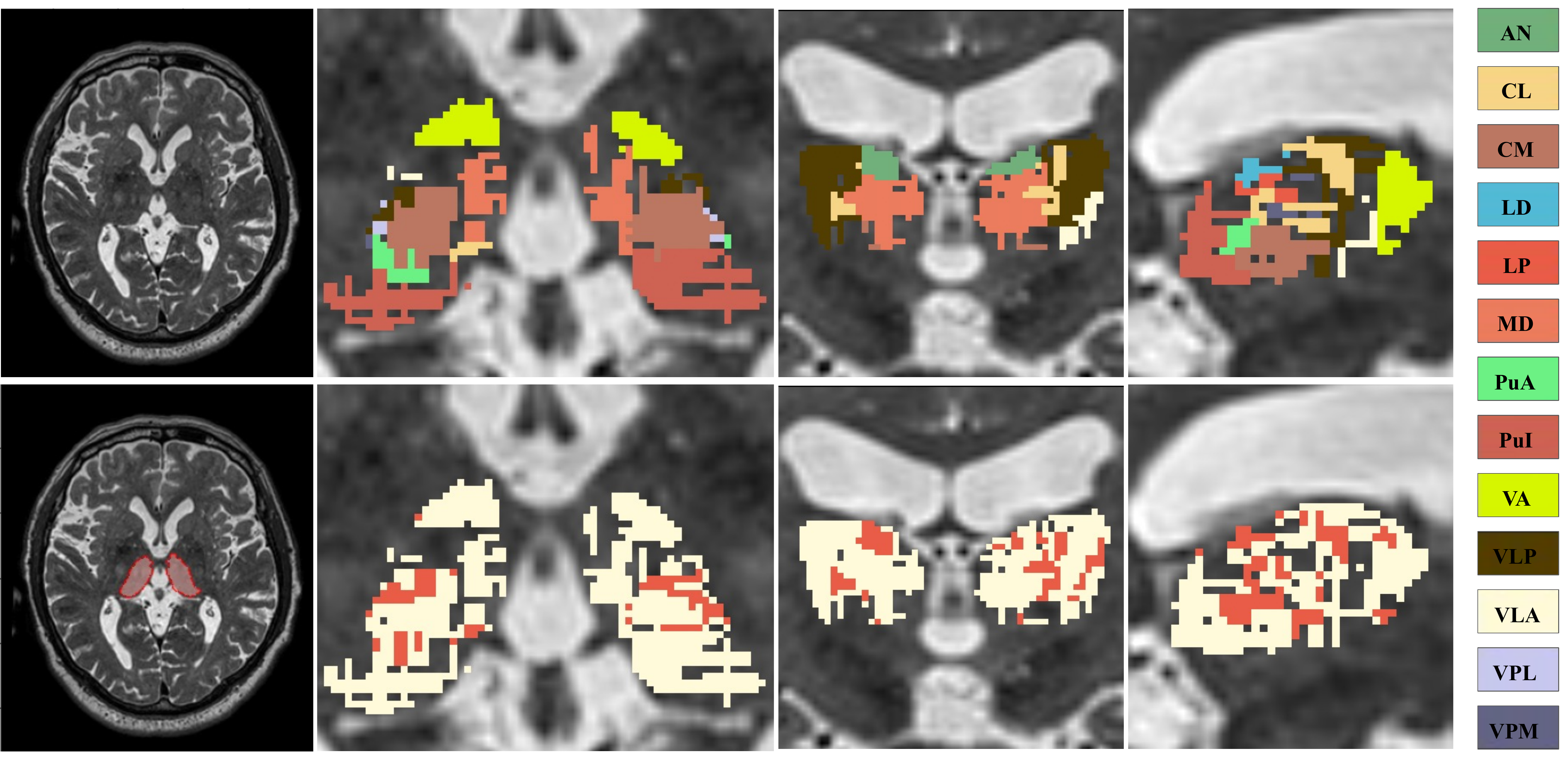} 
\caption{A 3DT2 weighted MR image and its corresponding thalamus mask are shown in the first column.
The next three columns are zoomed axial, coronal, and sagittal views, respectively. The top row has the thalamic nuclei labels (color key in the right-most column). The bottom row shows voxels with one label~(79.8\%) in white and multiple labels~(20.2\%) in red.}
\label{fig::visualize_data}
\end{figure}

Machine learning has proven successful at magnetic resonance~(MR) based thalamic nuclei parcellation~\cite{stough2014automatic, ziyan2006segmentation}. However, such methods rely on high-quality thalamic nuclei annotations for model training. This is a difficult task in practice for several reasons~\cite{iglesias2018probabilistic, niemann2000morel, su2019thalamus}: 1)~nuclei are small~\cite{keller2012volume}; 2)~low contrast between nuclei; 3)~delineations done on 2D slices while the nuclei are 3D structures, and 4)~independent labeling of separate nuclei can lead to gaps and overlaps---as shown in Fig.~\ref{fig::visualize_data}.

\begin{figure*}[!tb]
\centering
\includegraphics[width = 0.85\textwidth]{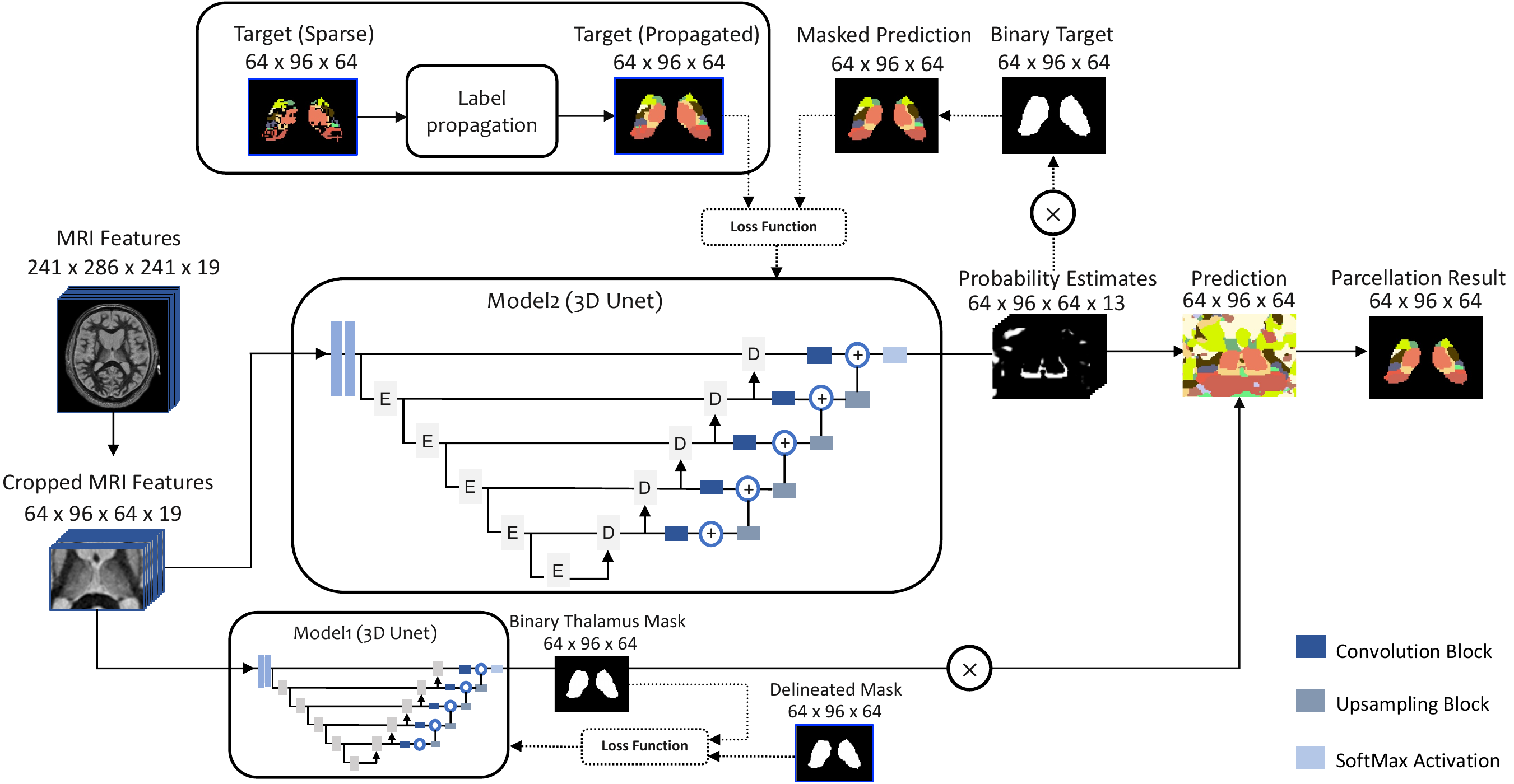}
\caption{An overview of the proposed model. The parcellation model takes 3D MRI volumes with 19 features as input and generates both a whole thalamus mask and thalamic nuclei labels. Two models are trained separately in a two-step way. The whole thalamus mask is applied to the thalamic nuclei labels to get the final parcellation result.} 
\label{fig::workflow}
\end{figure*}

To leverage imperfect annotations, efforts have been made towards training models with noisy labels. Gao~et~al.~\cite{gao2016segmentation} proposed a constrained segmentation propagation algorithm which uses information from sparse annotations and voxel-wise segmentation results from deep convolutional neural networks, and formulates it into a fully-connected conditional random field. However, the algorithm only works in 2D. Breve~et~al.~\cite{breve2019interactive} proposed a two-stage graph-based label propagation method, spreading ``scribbles'' to complete labels; this method may not be transferable to thalamus nuclei segmentation due to the low image contrast between nuclei. 

In this paper, we propose a robust thalamic nuclei segmentation framework that can be trained with noisy, incomplete annotations. Given the original annotation, which may contain gaps and have multiple labels assigned to voxels, we remove all overlapping labels and perform a label propagation using the random walk algorithm on an edge map. The propagated whole thalamus mask and thalamic nuclei labels train a two-step parcellation network based on 3D Unet. The class imbalance issue due to the small nuclei sizes is mitigated by focusing only on the thalamus. Compared to Yan~et~al.~\cite{chang2023spie} and a baseline method trained without label propagation, our proposed method achieves both quantitatively and qualitatively better performance.

\section{Methodology}
\label{sec:method}
An overview of our thalamic nuclei parcellation framework is shown in Fig.~\ref{fig::workflow}; the following subsections provide a detailed description.
%
% This felt repeatative to me. -AJC
%
% To mitigate the small size of thalamic nuclei, we use a 2-step model where the thalamic nuclei maps are masked by the whole thalamus mask so that the model can neglect the dominating background voxels during training. We remove conflicting overlapping labels and then do label propagation to fill in missing labels.
%

\subsection{Label Propagation}
\label{sec:label_propagation}
\begin{figure}[!tb]
\centering
\includegraphics[width=1\columnwidth]{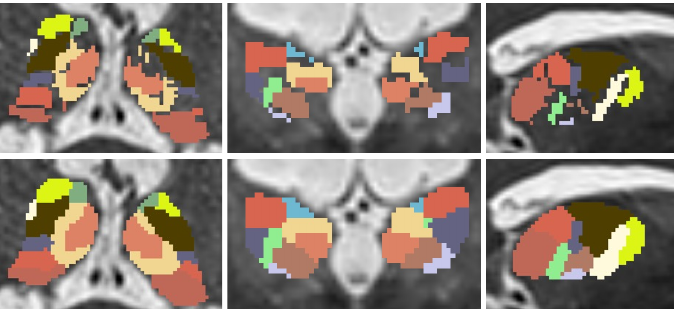}
\caption{Label propagation results. Top: Original noisy labels overlaid on 3DT2 weighted. Bottom: Propagated labels overlaid on 3DT2 weighted. Left to right are axial, coronal, and sagittal views. All the images come from the same subject.} 
\label{fig::label_propagation}
\end{figure}

As input, a manual delineation has thalamic voxels with no label, one label, or multiple labels.  We start by setting all voxels with multiple labels as unlabeled.  We next use the random walk algorithm~\cite{grady2006random} to propagate the remaining labels into unlabeled voxels within the thalamus, where labeled voxels are kept fixed. The random walk algorithm takes an input image, such as a magnetization-prepared rapid gradient-echo~(MPRAGE) MR sequence, and forms the undirected graph $G = (N, E)$. The nodes, $N$, are the voxels of the image and the edges, $E$, denote the connection between two neighboring nodes, in this case the six neighbors of a voxel. The edge, $e_{ij}$, between nodes $i$ and $j$ has weight:
\begin{equation*}
   w_{ij} = \exp\big( - \beta |g_{i}-g_{j}|^{2}  \big),
\end{equation*}
where $g$'s are the intensity values of nodes ($i$ or $j$) and $\beta$ is a hyperparameter. In our experiments, we will compare results that use intensities arising from different images: T1-map, MPRAGE, FGATIR~\cite{sudhyadhom2009ni, tohidi2023spie}, and T2-weighted.  Label propagation is formulated as a combinatorial Dirichlet problem~\cite{biggs1997algebraic} with desired random walker probabilities.
% The Dirichlet problem is defined as finding a harmonic function subject to the boundary value and minimizing Dirichlet Integral; the formula of Dirichlet Integral is:
%
% \begin{equation}
%   D[u] = \frac{1}{2} \int_{\Omega} |\bigtriangledown u |^2 d\Omega ,
% \end{equation} 
% and the harmonic function is defined as:
% \begin{equation}
%  \bigtriangledown^2 u =0 ,
% \end{equation} 
% where $u$ is a field and $\Omega$ is a region. For the label propagation problem, $x$ is defined as the random field over $\{x_1, .., x_i,..x_n \}$, where $x_i\in x$ and $i\in N$. $x_i$ represents the probabilities of node $i$ belonging to each label in the label set $\{l_1, .., l_m \}$.
The label propagation problem based on Dirichlet integral can be achieved by:
\begin{equation*}
   \argmin_{\bm{X}} \frac{1}{2} \sum_{e_{ij}\in E} w_{ij}(\bm{x}_{i} - \bm{x}_{j})^2 ,
\end{equation*}
where $\bm{X}$ is a random field over $\{\bm{x}_1, \ldots, \bm{x}_n \}$, with $\bm{x}_i$ representing the probability that node $i$ belongs to the label set $\{l_1, \ldots, l_m\}$. After label propagation, the refined labels can be used to train the model introduced in Sec.~\ref{sec:segmentation_model} without needing hard label conversion. Qualitative results of label propagation based on random walk are shown in Fig.~\ref{fig::label_propagation}.

\subsection{Thalamus Nuclei Parcellation}
\label{sec:segmentation_model}
We first use a 3D Unet~\cite{cciccek20163d} to segment the whole thalamus, outputting a whole thalamus mask~\cite{shao2022evaluating}. We then use a second 3D Unet to identify 13 classes corresponding to the 13 thalamic nuclei labels in each thalamic hemisphere. These 13 class predictions are then multiplied by the predicted whole thalamus mask to get the final thalamic nuclei masks.

During the training of the second network, we only calculate the loss inside the thalamus to avoid the class imbalance issue that would otherwise arise from background voxels. For comparison, we trained the second network in two ways.  As a baseline method, we trained the network using only the original annotations.  Since these annotation contain voxels with multiple labels, we converted the labels into membership vectors where a voxel with $n$ labels assigns $1/n$ membership to each label. The network was trained with cross-entropy to produce 13 thalamic nuclei class labels.
We also trained the second network using the propagated labels and a Dice loss.

\begin{table*}[!tb]
\caption{Dice score of the 10-fold cross-validation experiment. The ``Overall'' label is a volume weighted combination of the thirteen Dice scores. Dice score is expressed as a percentage. \textbf{Key:} Yan et al. - A comparison method that uses UMAP to reduce 19D data features (either to 2D, 3D, or 4D), followed by KNN to perform voxel-wise classifications; Unet - Unet with noisy labels; UnetT1 - Unet with propagated labels using T1 map; UnetMP - Unet with propagated labels using MPRAGE; UnetFG - Unet with propagated labels using FGATIR; UnetT2w - Unet with propagated labels using 3D T2 weighted images.}
\label{tb:comparison}
\centering
\resizebox{0.95\textwidth}{!}{
\begin{tabular}{l c c c c c c c}
\toprule
\multicolumn{1}{c}{\textbf{Method}}  &\textbf{Overall}  &\textbf{AN} &\textbf{CL}
&\textbf{CM} &\textbf{LD} &\textbf{LP} &\textbf{MD} \\
\cmidrule(lr){1-8}
\textbf{Yan et al.} &$60.33\pm4.06$ &$21.08\pm10.06$ &$32.37\pm6.78$
&$32.85\pm5.12$ &$0.74\pm0.08$ &$46.20\pm5.54$ &$72.81\pm3.60$ \\
\textbf{Unet} &$87.10\pm8.19$ &$82.24\pm11.94$ &$68.94\pm10.56$
&$78.48\pm10.56$ &$52.34\pm21.92$ &$86.28\pm8.02$ &$92.91\pm6.70$ \\
\textbf{UnetT1} &$88.70\pm7.56$ &$85.99\pm9.07$ &$70.25\pm12.99$
&\bm{$85.79\pm8.75$} &$57.87\pm24.35$ &$88.42\pm7.93$
&\bm{$93.45\pm4.51$} \\
\textbf{UnetMP} &$\textbf{88.95}\pm\textbf{6.37}$
&$\textbf{89.32}\pm\textbf{7.50}$ &$71.33\pm10.25$ &$84.08\pm8.38$
&$\textbf{58.40}\pm\textbf{31.93}$ &\bm{$89.79\pm5.68$}
&$93.44\pm4.87$ \\
\textbf{UnetFG} &$88.89\pm5.86$ &$88.80\pm6.22$ &$70.78\pm9.26$
&$85.19\pm4.69$ &$55.11\pm31.19$ &$86.96\pm8.57$ &$93.32\pm4.03$ \\
\textbf{UnetT2w} &$87.87\pm7.18$ &$88.28\pm10.55$
&\bm{$71.98\pm9.25$} &$84.26\pm6.78$ &$41.08\pm34.83$
&$86.60\pm7.83$ &$92.61\pm3.85$ \\
\cmidrule{1-8}
\multicolumn{1}{c}{\textbf{Method}} &\textbf{PuA} &\textbf{PuI} &\textbf{VA} &\textbf{VLP} &\textbf{VLA} &\textbf{VPL} &\textbf{VPM}\\
\cmidrule(lr){1-8}
\textbf{Yan et al.} &$8.29\pm6.10$ &$86.82\pm2.27$ &$66.22\pm4.24$ &$61.98\pm3.10$ &$15.45\pm6.95$ &$37.79\pm5.07$& $35.66\pm6.42$\\
\textbf{Unet} &$48.78\pm20.65$ &$97.31\pm2.74$ &$89.26\pm13.97$ &$89.44\pm6.59$ &\bm{$72.38\pm17.54$} &$77.44\pm12.96$ &$66.23\pm23.94$\\
\textbf{UnetT1} &$57.49\pm19.83$ &$97.67\pm1.99$ &$93.10\pm4.67$ &$89.92\pm9.69$ &$68.58\pm28.92$ &$81.14\pm9.81$ &$71.93\pm18.88$\\
\textbf{UnetMP} &$54.42\pm18.58$ &$97.60\pm2.01$ &$93.96\pm4.31$ &$92.41\pm4.01$ &$53.01\pm33.34$ &$81.19\pm7.86$ &$67.25\pm27.73$\\
\textbf{UnetFG} &$50.76\pm20.53$ &\bm{$97.81\pm2.12$} &$94.83\pm3.92$ &\bm{$92.51\pm3.40$} &$50.83\pm35.80$ &\bm{$81.55\pm8.44$} &$75.25\pm13.11$\\
\textbf{UnetT2w} &\bm{$58.46\pm14.94$} &$97.60\pm1.98$ &\bm{$95.09\pm4.01$} &$88.88\pm9.47$ &$46.02\pm34.03$ &$79.33\pm11.55$ &\bm{$78.91\pm13.59$}\\
\bottomrule  
\end{tabular}}
\end{table*}

\begin{figure*}[!t]
\centering
\includegraphics[width = 0.99\textwidth]{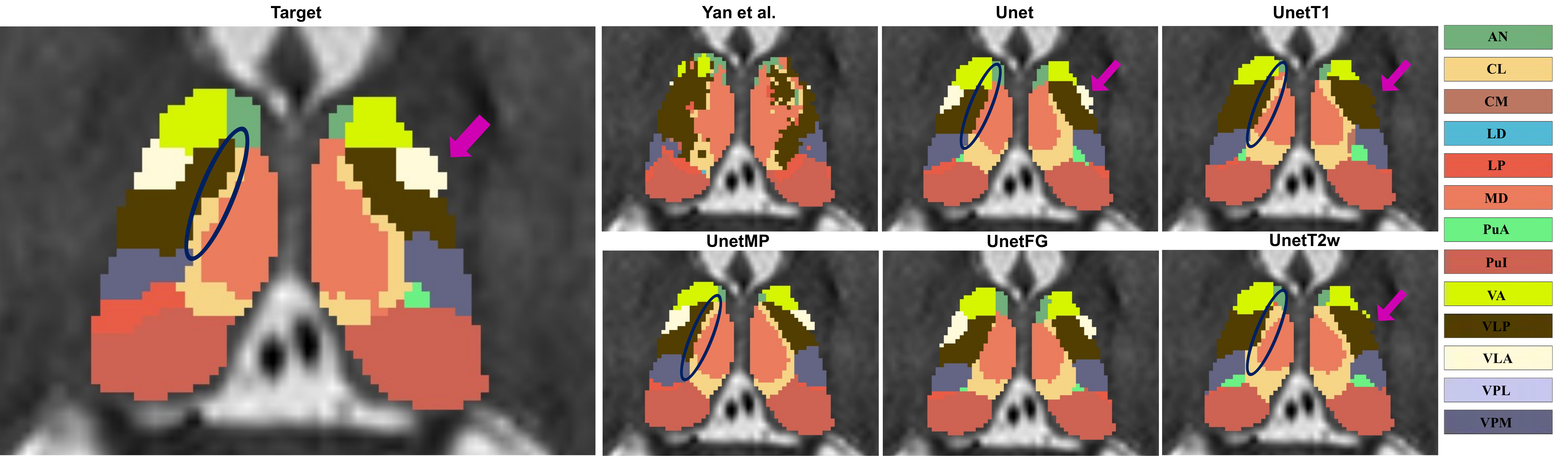}
\caption{Qualitative comparison of different parcellation results. The target after label propagation and parcellation results are shown respectively. The target is a label fusion result of the propagated labels calculated based on T1 map, MPRAGE, FGATIR, and 3DT2 weighted using majority voting implemented from:  \texttt{https://github.com/FETS-AI/LabelFusion}.} 
\label{fig::visualize results}
\end{figure*}

\section{Experiments and Results}
\noindent\textbf{Dataset}\qquad
For both training and testing, we used MR scans from 30 subjects. For each subject, multi-contrast MR features, including fast gray matter acquisition T1 inversion recovery~(FGATIR)~\cite{sudhyadhom2009ni}, MPRAGE, three-dimensional T2-weighted~(3D-T2w), T1 map, and diffusion tensor imaging~(DTI) were available.
%
% AN - Anterior Nucleus; CL - Central Lateral Nucleus; CM - Center Median Nucleus; LD - Lateral Dorsal Nucleus; LP - Lateral Posterior Nucleus; MD - Mediodorsal; PuA - Anterior Pulvinar; PuI - Inferior Pulvinar - VA - Ventral Anterior Nucleus; VLA - Ventral lateral Anterior Nucleus; VLP - Ventral Lateral Posterior Nucleus; VPL - Ventral Posterior Lateral Nucleus; VPM - Ventral Posterior Medial Nucleus.
%
The DTI data were used to generate diffusion measures, including fractional anisotropy~(FA), mean diffusivity~(MD), axial diffusivity~(AD), radial diffusivity~(RD), trace~(Tr), 3 Westin Indices~(3WI), mode, eigenvectors, and eigenvalues. Eigenvectors and eigenvalues were further used to construct the 5D Knutsson vector and edge map \cite{knutsson1985producing}. All images were registered to the MPRAGE, non-directional images were normalized to the range of $[0, 1]$, while directional features were normalized to $[-1, 1]$. We also perform center-cropping to crop the images from $241 \times 286 \times 241$ to $64 \times 96 \times 64$. The MPRAGE, FGATIR, 3D-T2, T1 map, FA, MD, AD, RD, Tr, 3WI, mode, and the 5D Knutsson vector and edge map were combined into a 19-dimensional feature vector. 

The 30 subjects were semi-automatically annotated using the following steps: 1)~the Morel atlas~\cite{niemann2000morel} was affinely registered to a subjects MPRAGE image; 2)~the atlas labels were individually transformed into the subjects space, which can lead to discontinuities and label overlaps; 3)~atlas labels were reviewed and manually corrected, this correction (as shown in Fig.~\ref{fig::visualize_data}) does not resolve the label gaps and overlaps. The nuclei used in this work are the: anterior nucleus~(AN), central lateral~(CL), center median~(CM), lateral dorsal~(LD), lateral posterior~(LP), mediodorsal~(MD), anterior pulvinar~(PuA), inferior pulvinar~(PuI), ventral anterior~(VA), ventral lateral anterior~(VLA), ventral lateral posterior~(VLP), ventral posterior lateral~(VPL),ventral posterior medial~(VPM).
%

% AN - Anterior Nucleus; CL - Central Lateral Nucleus; CM - Center Median Nucleus; LD - Lateral Dorsal Nucleus; LP - Lateral Posterior Nucleus; MD - Mediodorsal; PuA - Anterior Pulvinar; PuI - Inferior Pulvinar - VA - Ventral Anterior Nucleus; VLA - Ventral lateral Anterior Nucleus; VLP - Ventral Lateral Posterior Nucleus; VPL - Ventral Posterior Lateral Nucleus; VPM - Ventral Posterior Medial Nucleus.
%

Our baseline Unet method refers to Unet trained with the original (noisy) labels. We also trained four more Unets using propagated labels from the random walk algorithm where $\beta$ is $10,000$ and used the T1 map, MPRAGE, FGATIR, and 3D-T2w images as the edgemap. For this, we did a 10-fold cross validation with 25 training, 2 validation, and 3 testing subjects in each fold. For the Unet training, we used the SGD optimizer with an initial learning rate of $5 \times 10^{-2}$, which decreased by $10\%$ every $20$ epochs, and a momentum of $0.9$. The models were trained for $200$ epochs. During evaluation, we excluded voxels with ambiguous annotations on the testing data and calculate the Dice coefficient of the prediction multiplied by the binary version of target. We calculated Dice for each thalamus nuclei class and report a volume-weighted mean of the Dice score for each of the 13 classes (Table~\ref{tb:comparison}).

The results in Table~\ref{tb:comparison} show clear improvement over the previously reported method of Yan~et~al.~\cite{chang2023spie}. The Yan method used the same 19 features that we use here and then applied the UMAP~\cite{mcinnes2018oss} method followed by $k$ nearest neighbors~(KNN) in the resulting 2D space.  Among the Unet methods, we observe overall best performance with the propagated labels from the MPRAGE image. Individual nuclei may favor a particular contrast, but these differences are mostly small. Only the VLA shows better results when using the baseline UNet.  

Qualitative thalamic nuclei parcellation results are shown in Fig.~\ref{fig::visualize results}. We observe that each variation of our proposed method achieves better performance than Yan~et~al.~\cite{chang2023spie}.
Most nuclei segmented using models with propagated labels show higher concordance with the target delineation than the baseline method. For example, the central lateral nucleus (CL) class, as highlighted by the blue circle, is clearly predicted by UnetT1, UnetT2w and UnetMP, while the baseline Unet predicts it as MD. However, the ventral lateral anterior nucleus~(VLA) class, as highlighted by the purple arrows, appears visually better in the baseline Unet than UnetT1 and UnetT2w, both of which predict the VLA class as VLP.

% Why are you looking at the VLA class?
%
%However, while most Unet-segmented nuclei show high concordance with the target delineation, the ventral lateral anterior nucleus~(VLA) class, as highlighted by the purple arrows, appears visually best in the baseline Unet. In this example, both UnetT1 and UnetT2w predicted the VLA class as VLP whereas UnetMP and UnetFG got better results.

% The results of the Dice score are shown in Table~\ref{tb:comparison}. The performance of the baseline U-net trained on the original (noisy) labels had an overall Dice score of 87.10\%. Among the methods of Unet trained with propagated labels, when using the T1 map to calculate edge map, the overall dice score was improved to 88.70\%; when using the MPRAGE to calculate edge map, the overall dice score was improved to 88.95\%; when using the FGATIR to calculate edge map, the overall dice score was improved to 88.89\%; when using the 3D T2 weighted to calculate edge map, the overall dice score was improved to 87.87\%.

\section{Conclusion}
\label{sec:conclusion}
In this paper, we proposed a thalamic nuclei parcellation model which can be trained with noisy annotations. A random walk based label propagation was used to refine the original annotations with conflicting and missing voxels. Compared with existing models and a baseline Unet model trained with the original labels, our proposed model trained with our propagated labels achieved better performance, both quantitatively and qualitatively. We believe our proposed model can be extended to other medical imaging applications where noisy annotations are common.

\section{Acknowledgments}
\label{sec:acknowledgments}
This work was supported in part by the NIH through the National Institute of Neurological Disorders and Stroke~(NINDS) grant R01-NS105503 (PI: R.P.~Gullapalli). The data was acquired in line with the principles of the Declaration of Helsinki. Approval was granted by an IRB Committee of the University of Maryland School of Medicine.
%
% IRB Approval IDs would be nice.
%

\bibliographystyle{IEEEbib}
\bibliography{Main}
\end{document}